\documentclass[]{raa}            
\usepackage{graphicx,times}
\usepackage{natbib}
\usepackage{graphicx}
\usepackage{epstopdf}
\usepackage{color}

\providecommand{\bm}[1]{\mbox{\boldmath$#1$\unboldmath}}

\begin{document}

   \title{Quasars in the Galactic Anti-Center Area from the LAMOST DR3}

 \volnopage{ {\bf 2017} Vol.\ {\bf X} No. {\bf XX}, 000--000}
   \setcounter{page}{1}

   \author{Zhi-Ying Huo
      \inst{1}
   \and Xiao-Wei Liu
      \inst{2,3}
   \and Jian-Rong Shi
      \inst{1}    
    \and Mao-Sheng Xiang
      \inst{1}    
  \and Yang Huang
      \inst{2}
  \and Hai-Bo Yuan
      \inst{4}
  \and Jian-Nan Zhang
      \inst{1}    
 \and Wei Zhang
      \inst{1}
 \and Jian-Ling Wang
      \inst{1}
 \and Yu-Zhong Wu
      \inst{1}  
  \and Zi-Huang Cao
      \inst{1}
   \and Yong Zhang
      \inst{5}
  \and Yong-Hui Hou
      \inst{5} 
  \and Yue-Fei Wang
      \inst{5}    
   }


   \institute{National Astronomical Observatories, Chinese Academy of Sciences,
                  Beijing 100012, P.R. China; {\it Email: zhiyinghuo@bao.ac.cn}  
        \and   Department of Astronomy, Peking University,
                  Beijing 100871, P.R. China
        \and    Kavli Institute for Astronomy and Astrophysics, Peking University,
                  Beijing 100871, P.R. China    
       \and    Department of Astronomy, Beijing Normal University, Beijing, 100875, P.R. China           
       \and    Nanjing Institute of Astronomical Optics \& Technology, 
                  Chinese Academy of Sciences, Nanjing 210042, P.R. China 
\\
\vs \no
   {\small Received [2016] [Dec] [9]; accepted [2017] [Feb] [5] }
}

\abstract{
We present a sample of quasars discovered in the area of Galactic Anti-Center (GAC) of
$150^{\circ} \leq l \leq 210^{\circ}$ and $|b| \leq 30^{\circ}$, 
based on the LAMOST Data Release 3 (DR3). The sample contains 151 spectroscopically confirmed quasars.
Among them 80 are newly discovered with the LAMOST. 
All those quasars are very bright, with $i$ magnitudes peaking around 17.5\, mag.
All the newly quasars are discovered serendipitously, targeted originally with the LAMOST as stars
of bluer colours,
except for a few targeted as variable, young stellar object candidates. 
This bright quasar sample at low Galactic latitudes will help fill the gap in the spatial distribution of 
known quasars near the Galactic disk that are used to construct astrometric reference frame for the purpose of
accurate proper motion measurements, for example, Gaia. 
They are also excellent tracers to probe the kinematics and chemistry of the interstellar medium of the Milky Way disk
and halo via absorption line spectroscopy.
}

   \authorrunning{Z.~Y. Huo et al.}            
   \titlerunning{Quasars in the Galactic Anti-Center Area from the LAMOST DR3} 
   \maketitle


%
%

\section{Why Searching for Quasars in the Galactic Anti-Center Area?}

Until the recent twelfth data release Quasar Catalog, the number of spectroscopically confirmed 
quasars has reached $\sim$400,000 
(Schneider et al. 2010; P{\^a}ris et al. 2012, 2014, 2016; Souchay et al. 2015 and reference therein).
Most of the quasars are identified with the 
Sloan Digital Sky Survey (SDSS, York et al. 2000), and their spatial distribution is largely homogeneous
in the approximately 7,500 deg$^2$ around the North Galactic Cap and about 3,100 deg$^2$ 
surrounding the South Galactic Cap (see Schneider et al. 2010; P{\^a}ris et al. 2012, 2014, 2016). 
However, near the Galactic disk, especially for $|b| \leq 30^{\circ}$,  
the number of spectroscopically confirmed quasars remains very small,  only a few bright ones have
been discovered hitherto, due to the lack of systematic spectroscopic surveys in those low Galactic
latitude, often highly dust extinct Galactic disk regions (see Souchay et al. 2015, 
V\' {e}ron-Cetty \& V\' {e}ron 2010 and reference therein).

Distant quasars are ideal objects to construct inertial reference frame for accurate proper motion 
measurements by all sky astrometric survey mission such as Gaia 
(Perryman et al. 2001; Gaia collaboration, Brown, et al. 2016; Gaia collaboration, Prusti, et al. 2016; 
Lindegren et al. 2016). Proper motion measurements are of vital importance for the Galactic studies. 
Gaia uses a large number and all-sky distributed quasars, based on Souchay et al. (2008), 
builds an optical materialization of the International Celestial Reference System 
(Andrei. et al. 2009). However, the number of available quasars near the Galactic plane is quite low.
Bright, distant quasars are also excellent tracers to study the foreground
interstellar/intergalactic medium (ISM/IGM) along the line of sight and probe their properties
 such as the distribution, chemical composition, and kinematics.
For instance, based on quasar absorption line spectroscopy, Savage et al. (2000) and 
Schneider et al. (1993) show that the halo gas of the Milky Way processes a wide range of ionization states, 
chemical compositions, and kinematics.

The LAMOST Spectroscopic Survey of the Galactic Anti-center (LSS-GAC), 
is a major component of the LAMOST Galactic surveys (Liu et al. 2014; Yuan et al. 2015), 
aiming to survey a significant volume of the Galactic thin/thick disks and halo for a contiguous 
sky area of over 3400 deg$^2$ centered on the GAC, within $150^{\circ} \leq l \leq 210^{\circ}$ and 
$|b| \leq 30^{\circ}$. 
LSS-GAC targets are selected from the Xuyi Schmidt Telescope Photometric Survey of the 
Galactic Anti-centre (XSTPS-GAC; Liu et al. 2014; Zhang et al. 2014) photometric catalogs.
The basic idea of LSS-GAC target selection is to uniformly and randomly select stars from 
the colour-magnitude diagrams using a Monte Carlo method. Stars of extremely blue or red
colours are preferentially selected and targeted, see Yuan et al. (2015) for more detail. 
As a consequence, one expects that some targets targeted by the LSS-GAC should contain
a number of quasar candidates.


\section{Results} 

The current work is based on the LAMOST Data Release 3 (DR3), which includes data collected
from October 24, 2011 (the beginning of the Pilot Survey) to May 30, 2015. The quasar classifications
and redshift measurements are produced by the LAMOST 1D pipeline.
In the LAMOST 1D pipeline, the observed spectra are cross-correlated with stellar, galaxy, and quasar 
template library, four spectral type classifications, namely STAR, GALAXY, QUASAR 
and UNKNOWN are produced, radial velocity for stars or redshift for galaxies and quasars are 
measured simultaneously (see Luo et al. 2015 for more details).  However, 
all the 1D pipeline identifications and redshifts for quasars and galaxies are visually inspected and 
adjusted if necessary before the final data release.

In this letter, we present 151 unique quasars from the LAMOST DR3. Amongst them 80 are reported here for the 
first time. The remaining 71 are listed in the Large Quasar Astrometric Catalog (LQAC) compiled recently 
by Souchay et al. (2015). 
Of the 151 quasars, 143 are identified by the LAMOST pipeline with classification as `QUASAR' (Luo et al. 2015). 
In addition, we have visually examined approximately 80,000 spectra classified as `UNKNOWN' 
by the LAMOST pipeline, either because of
low signal noise ratios (SNRs) or for other reasons, of targets near the Galactic disk, 
of $|b| \leq 15^{\circ}$. The effort is to identify as many quasars as possible, even from spectra of
relatively low SNRs that the pipeline may fail to work properly. 
Quasars are easily identified given their characteristic broad emission line spectra, 
H$\alpha$ $\lambda$6563, [O~{\sc iii}] $\lambda\lambda$4959, 5007, H$\beta$ $\lambda$4861, 
Mg~{\sc ii} $\lambda$2800, 
C~{\sc iii}] $\lambda$1908, C~{\sc iv} $\lambda$1549 and Ly$\alpha$ $\lambda$1216 lines are shifts 
into the LAMOST spectra wavelength coverages with quasar redshift increasing.
Finally, we have identified 8 quasars from those `UNKNOWN' spectra.
For the example of LAMOST quasar spectra, please refer to our series of papers about background
quasars in the vicinity of M\,31 and M\,33 (Huo et al. 2010, 2013, 2015). 
All the spectra have not been flux-calibrated, we will not show them here.

Table 1 lists the information of all the 151 quasars in the GAC area identified with the LAMOST, including
target name, J2000 coordinates of RA and Dec in decimal degree, redshift, the observed 
$g$, $r$, $i$ magnitude if available, magnitude type (`magtype'), object type from the LAMOST 
input catalog (`objtype'), identified lines and notes. 
In the last Column of Table 1, a flag `N' indicates that the quasar is
newly discovered by the LAMOST, whereas `L' indicates it is listed in LQAC (Souchay et al. 2015).
None of the 151 quasars were originally targeted by LAMOST as quasar candidates. 
Amongst them, 141, 1 and 9  objects were originally targeted as stars of blue colours, 
variable sources selected from the WISE (Wright et al. 2010) photometric catalog, 
and young stellar objects (`yso'), respectively. The distributions of $i$ band magnitudes of this
sample of quasars peaks at 17.5\, mag, with
most them having $i$ magnitude between 16 and 18. The majority (141/151) of the quasars,  
have a redshift lower than 2.2. Only 11 quasars have higher redshifts, with the highest redshift 
value being 3.43.

Figure 1 plots the spatial distribution of this sample of quasars found in the GAC area,
in Galactic coordinates, 
including the 80 quasars newly identified with the LAMOST and the 71 previously known ones. 
The newly discovered quasars are distributed closer to the Galactic plan
than those previously known ones. This sample of quasars will help fill the gap in the
spatial distribution of quasars near the Galactic disk that are used to build 
astrometric reference frame for proper motion measurements. They are also valuable sources to
probe kinematics and chemistry of the ISM of the Milky Way disk and halo.

\begin{figure}[h!!!]
\centering
\includegraphics[width=9.0cm,angle=0]{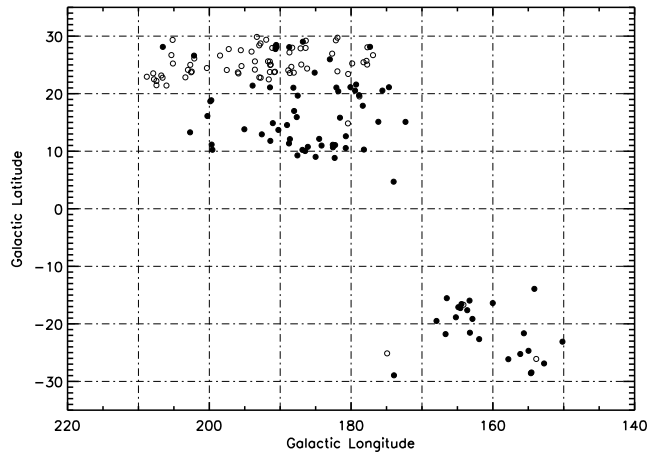}
\caption{Spatial distributions of all background quasars in the GAC area identified from the
 LAMOST DR3.
Filled circles represent newly discovered quasars, and open circles represent those already
listed in LQAC (Souchay et al. 2015).}
\label{qsocand}
\end{figure}

\section{Summary}
We present 151 quasars discovered in the GAC area of $150^{\circ} \leq l \leq 210^{\circ}$ and $|b| \leq 30^{\circ}$, 
from the LAMOST DR3, among them 80 are newly discovered with the LAMOST. 
These bright quasar will help fill the gap in the spatial distribution of known quasars near the Galactic disk
that are used to construct astrometric reference frame for the purpose of accurate proper motion measurements,
and will also served as tracers to probe the properties of the ISM of the Milky Way disk and halo.
We will presented more quasars near the Galactic disk in the future with more data collected by the LAMOST.

\begin{table}[h!!!]
\bc
\begin{minipage}[]{150mm}
\caption[]{Catalog of Quasars in the GAC area from the LAMOST DR3}
\label{tab_cat}
\end{minipage}
\setlength{\tabcolsep}{2pt}
\small
 \begin{tabular}{ccccccccccc}
  \hline\noalign{\smallskip}
Object  & R.A. & Dec. (J2000) & Redshift  & $g$ & $r$ & $i$ & magtype & objtype & lines & note$^*$ \\  
  \hline\noalign{\smallskip}

 J025226.35+331721.0    &  43.109796   &   33.289167  &  0.53   &  17.15   &  17.10  &   16.96   &  gri    &    Star   &  Mg~{\sc ii}, H$\gamma$, H$\beta$, [O~{\sc iii}] & N    \\ 
 J025311.52+285016.0    &  43.298027   &   28.837802  &  0.56   &  17.86   &  17.74  &   17.57   &  gri    &    Star   &  Mg~{\sc ii}, H$\beta$ & N    \\ 
 J025539.40+263218.3    &  43.914171   &   26.538418  &  0.53   &  17.78   &  18.29  &   17.75   &  gri    &    Star   &  Mg~{\sc ii}, [O~{\sc ii}], H$\gamma$, H$\beta$, [O~{\sc iii}] & N    \\ 
 J025540.49+264119.3    &  43.918725   &   26.688717  &  0.29   &  17.55   &  17.54  &   17.27   &  gri    &    Star   &  H$\gamma$, H$\beta$, [O~{\sc iii}], H$\alpha$ & N    \\ 
J025855.16+290001.3    &  44.729862   &   29.000370  &  2.48   &      -       &  -         &    17.30  &     i    &     yso     &  Ly$\alpha$, N~{\sc v}, Si~{\sc iv}, C~{\sc iv}, C~{\sc iii}] & L    \\ 
 J030613.11+294029.0    &  46.554666   &   29.674743  &  0.89   &  18.23   &  17.70  &   17.88   &  gri   &     Star   &  Mg~{\sc ii}, H$\gamma$ & N    \\ 
 J030850.72+283738.2    &  47.211350   &   28.627296  &  2.09   &  18.15   &  17.79  &   17.65   &  gri   &     Star   &  Ly$\alpha$, N~{\sc v}, Si~{\sc iv}, C~{\sc iv}, C~{\sc iii}], Mg~{\sc ii} & N    \\ 
 J031212.67+270143.7    &  48.052832   &   27.028832  &  3.43   &  18.49   &  17.93  &   17.71   &  gri   &     Star   &  O~{\sc vi}, Ly$\alpha$, N~{\sc v}, Si~{\sc iv}, C~{\sc iv}, C~{\sc iii}] & N    \\ 
 J031624.33+315225.0    &  49.101390   &   31.873619  &  0.17   &  18.24   &  17.56  &   16.69   &  gri   &     Star   &  H$\beta$, O~{\sc iii}, H$\alpha$ & N    \\ 
 J033243.77+385859.5    &  53.182404   &   38.983215  &  1.84   &  17.97   &  17.71  &  17.39    &  gri   &     Star   &  C~{\sc iv}, C~{\sc iii}], Mg~{\sc ii} & N\\

   \noalign{\smallskip}\hline
\end{tabular}
\ec
\tablecomments{1.0\textwidth}{$^*$N represents the quasar is newly discovered by the LAMOST, 
whereas `L' represents it is listed in LQAC (Souchay et al. 2015).
Only a portion of Table is shown here for illustration. The whole Table
contains information of 151 quasars is available in the online electronic version.}
\end{table}

\normalem
\begin{acknowledgements}

This work is supported by National Key Basic Research Program of China 2014CB845705, 
NSFC grant 11403038, 11473001 and U1531244. 
The Guoshoujing Telescope (the Large Sky Area Multi-Object Fiber Spectroscopic
Telescope; LAMOST) is a National Major Scientific Project built by the Chinese
Academy of Sciences. Funding for the project has been provided by the National
Development and Reform Commission. The LAMOST is operated and managed by the
National Astronomical Observatories, Chinese Academy of Sciences.  
 
\end{acknowledgements}



\begin{thebibliography}{99}
\small \setlength{\itemindent}{-3mm} \setlength{\itemsep}{-0.5mm}
\setlength{\baselineskip}{4.5mm}



\bibitem[Andrei et al.(2009)]{2009A&A...505..385A} Andrei, A.~H., Souchay, J., Zacharias, N., et al.\ 2009, \aap, 505, 385
\bibitem[Gaia Collaboration et al.(2016)]{2016arXiv160904172G} Gaia Collaboration, Brown, A.~G.~A., Vallenari, A., et al.\ 2016, arXiv:1609.04172
\bibitem[Gaia Collaboration(2016)]{2016arXiv160904153G} Gaia Collaboration,  Prusti, T., de Bruijne, J. H.J., et al. 2016, arXiv:1609.04153
\bibitem[Huo et al.(2010)]{2010RAA....10..612H} Huo, Z.-Y., Liu, X.-W., Yuan, H.-B., et al.\ 2010, Research in Astronomy and Astrophysics, 10, 612
\bibitem[Huo et al.(2013)]{2013AJ....145..159H} Huo, Z.-Y., Liu, X.-W., Xiang, M.-S., et al.\ 2013, \aj, 145, 159
\bibitem[Huo et al.(2015)]{2015RAA....15.1905H} Huo, Z.-X., Li, Y.-M., Li, X.-B., \& Zhou, J.-F.\ 2015, Research in Astronomy and Astrophysics, 15, 1905
\bibitem[Lindegren et al.(2016)]{2016arXiv160904303L} Lindegren, L., Lammers, U., Bastian, U., et al.\ 2016, arXiv:1609.04303
\bibitem[Liu et al.(2014)]{2014IAUS..298..310L} Liu, X.-W., Yuan, H.-B., Huo, Z.-Y., et al.\ 2014, Setting the scene for Gaia and LAMOST, 298, 310
\bibitem[Luo et al.(2015)]{2015RAA....15.1095L} Luo, A.-L., Zhao, Y.-H., Zhao, G., et al.\ 2015, Research in Astronomy and Astrophysics, 15, 1095
\bibitem[P{\^a}ris et al.(2012)]{2012A&A...548A..66P} P{\^a}ris, I., Petitjean, P., Aubourg, {\'E}., et al.\ 2012, \aap, 548, A66
\bibitem[P{\^a}ris et al.(2014)]{2014A&A...563A..54P} P{\^a}ris, I., Petitjean, P., Aubourg, {\'E}., et al.\ 2014, \aap, 563, A54
\bibitem[P{\^a}ris et al.(2016)]{2016arXiv160806483P} P{\^a}ris, I., Petitjean, P., Ross, N.~P., et al.\ 2016, arXiv:1608.06483 
\bibitem[Perryman et al.(2001)]{2001A&A...369..339P} Perryman, M.~A.~C., de Boer, K.~S., Gilmore, G., et al.\ 2001, \aap, 369, 339
\bibitem[Savage et al.(2000)]{2000ApJS..129..563S} Savage, B.~D., Wakker, B., Jannuzi, B.~T., et al.\ 2000, \apjs, 129, 563
\bibitem[Schneider et al.(1993)]{1993ApJS...87...45S} Schneider, D.~P., Hartig, G.~F., Jannuzi, B.~T., et al.\ 1993, \apjs, 87, 45
\bibitem[Schneider et al.(2010)]{2010AJ....139.2360S} Schneider, D.~P., Richards, G.~T., Hall, P.~B., et al.\ 2010, \aj, 139, 2360
\bibitem[Souchay et al.(2015)]{2015A&A...583A..75S} Souchay, J., Andrei, A.~H., Barache, C., et al.\ 2015, \aap, 583, A75
\bibitem[Souchay et al.(2008)]{2008A&A...485..299S} Souchay, J., Lambert, S.~B., Andrei, A.~H., et al.\ 2008, \aap, 485, 299
\bibitem[V{\'e}ron-Cetty \& V{\'e}ron(2010)]{2010A&A...518A..10V} V{\'e}ron-Cetty, M.-P., \& V{\'e}ron, P.\ 2010, \aap, 518, A10
\bibitem[Wright et al.(2010)]{2010AJ....140.1868W} Wright, E.~L., Eisenhardt, P.~R.~M., Mainzer, A.~K., et al.\ 2010, \aj, 140, 1868
\bibitem[York et al. 2000]{York2000} York, D.~G., et al.\ 2000, \aj, 120, 1579
\bibitem[Yuan et al.(2015)]{2015MNRAS.448..855Y} Yuan, H.-B., Liu, X.-W., Huo, Z.-Y., et al.\ 2015, \mnras, 448, 855
\bibitem[Zhang et al.(2014)]{2014RAA....14..456Z} Zhang, H.-H., Liu, X.-W., Yuan, H.-B., et al.\ 2014, Research in Astronomy and Astrophysics, 14, 456-470

\end{thebibliography}
\end{document}